\documentclass[11pt, letterpaper]{article}

\usepackage{arxiv}

\usepackage{times}
\usepackage[colorlinks=true, linkcolor=teal]{hyperref}
\usepackage{amsthm}
\usepackage{amsmath,amssymb,amsfonts}
\usepackage{helvet}
\usepackage{graphicx}
\usepackage{enumerate}
\usepackage{pgfplots}
\usepackage{mathtools,relsize}
\usepackage{float}
\usepackage{subfigure}
\usepackage{subcaption}
\usepackage{latexsym}
\usepackage{setspace}
\usepackage{algorithm,algorithmic}
\usepackage{graphicx}
\usepackage{textcomp}
\usepackage{xcolor}
\usepackage{multirow}
\usepackage{adjustbox}

\DeclareMathOperator*{\argmax}{arg\,max}

\begin{document}
%===============================================================
\title{User Acceptance Model for Smart Incentives in Sustainable Video Streaming towards 6G}

\author{
    Konstantinos Varsos \quad Adamantia Stamou \quad George D. Stamoulis \quad Vasillios A. Siris \\
     Department of Informatics, School of Information Sciences and Technology, \\
     Athens University of Economics and Business \and \\
     \texttt{\{kvarsos, stamouad, gstamoul, vsiris\}@aueb.gr} 
}

\date{}
%===============================================================

\maketitle
\begin{abstract}
The rapid growth of 5G video streaming is intensifying energy consumption across access, core, and data-center networks, underscoring the critical need for energy and carbon-efficient solutions. While reducing streaming bitrates improves energy efficiency, its success hinges on user acceptance—particularly when lower bitrates may be perceived as reduced quality of experience (QoE). Therefore, there is a need to develop transparent, user-centric incentive models that balance sustainability with perceived value. We propose a user-acceptance model that combines diverse environmental awareness, personalized responsiveness to incentives, and varying levels of altruism into a unified probabilistic framework. The model incorporates dynamic, individualized incentives that adapt over time. We further enhance the framework by incorporating (i) social well-being as a motivator for altruistic choices, (ii) provider-driven education strategies that gradually adjust user acceptance thresholds, and (iii) data-driven learning of user traits from historical offer–response interactions. Extensive synthetic-data experiments reveal the trade-offs between provider cost and network flexibility, showing that personalized incentives and gradual behavioral adaptation can advance sustainability targets without compromising stakeholder requirements.
\end{abstract}


\section{Introduction}\label{sec:Introduction}

The growth of 5G video streaming —especially high-definition and 4K content— has sharply increased energy consumption across access, core, and data-center networks \cite{Hossfeld2023AGE}. To mitigate the environmental impact of next-generation media services, 6G architectures must embed energy awareness by design. Promoting lower-bitrate streaming as a sustainable choice requires understanding user priorities and perceived QoE trade-offs \cite{Hossfeld2023AGE}, while incentive mechanisms--such as eco-friendly defaults, personalized rewards, and visible social impact--should reflect individual responsiveness and environmental awareness \cite{Damiani2013AnAB,Seger2023ReducingTI}. European research projects such as EXIGENCE \cite{EXIGENCE} and NUDGE \cite{nudge} reflect growing interest in behavioral interventions and policy-driven innovation for sustainable ICT services.

This work develops and evaluates a \emph{user–acceptance model} based on smart incentives, extending the economic-incentive framework of \cite{Krasopoulos2022FlexibilityMF}. The present model captures (i) heterogeneity in environmental awareness, (ii) personalized responsiveness to both economic and behavioral incentives, (iii) the impact of altruism for collective environmental welfare, and (iv) policies to improve responsiveness towards greener practices. In particular, our model focuses on enabling more sustainable streaming behaviours, extending prior demand–response formulations by utilizing the \emph{greenness factor} $\gamma_n$ from \cite{Hossfeld2023AGE}, which modulates each user’s QoE mapping, and introduces the personalized acceptance threshold $r_{\min,n}$ that represents the minimum incentive needed to offset the perceived loss in utility when switching from high to low video resolution. Further, the model incorporates stochastic acceptance through a sigmoid response as in \cite{Krasopoulos2022FlexibilityMF}, allowing the study of dynamic policies whereby a service provider offers individualized incentives and adapts them over time. Accordingly, the proposed model advances the state of the art through three novel extensions. First, we embed social well-being \cite{Minou2017TheEO} into user utility to capture altruistic behavior. Second, we assess how shifts in user preferences affect provider outcome. Furthermore, we analyze how the provider leverages past offer-response patterns to estimate user characteristics, to optimize the estimated user traits, and to optimize sustainability gains.

The remainder of this paper is organized as follows. Related work is presented in Section \ref{sec:related_work}. The proposed User Acceptance Model is analyzed in Section \ref{sec:user acceptance model}, followed by the aforementioned model extensions in Section \ref{sec:extensions}. 
Section \ref{sec:evaluation} presents a comprehensive evaluation using a synthetic data generator and numerical experiments to quantify the trade-off between provider cost and network flexibility. Finally, conclusions are provided in Section \ref{sec:conclusions}.

\section{Related Work}\label{sec:related_work}

Incentives are pivotal for promoting energy optimization and carbon reduction, enabling lasting change and can scale across users to deliver broad sustainability impact \cite{Seger2023ReducingTI}, \cite{Istrate2024}. Actually, well-designed behavioral cues can often outperform financial incentives \cite{Damiani2013AnAB}, empowering consumers through feedback, goal setting, and tailored information to foster collective awareness and mitigate rebound effects. Smart features such as real-time metering, neighbor comparisons, and personalized tips enhance engagement and support sustainable habits \cite{Damiani2013AnAB,Seger2023ReducingTI}. In fact, information provisioning combined with self-monitoring led to significant behavioral change, reducing carbon impact by over 19 per cent in users’ video streaming activities \cite{Seger2023ReducingTI}.

Higher video quality in streaming directly increases energy use and $\text{CO}_2$ emissions, creating a trade-off between user experience and environmental impact. Hossfeld et al. \cite{Hossfeld2023AGE} showed that small decreases in video quality--from "excellent" to "good"--can yield significant emission reductions. For that, they considered a “green user”, i.e., a user who values energy efficiency in their QoE assessment, with their willingness to reduce their video quality represented as a greenness factor $\gamma$. 

This paper extends the state of the art, as we treat $\gamma$ as a behavioral trait affecting the incentive mechanism and provider policies, allowing explicit modeling of user heterogeneity and learning. While \cite{Hossfeld2023AGE} embeds $\gamma$ solely in the QoE utility, we integrate it into a dynamic incentive framework where providers adapt offers based on user characteristics.

Compared with \cite{Krasopoulos2022FlexibilityMF}, which models demand-response participation through economic incentives and device-level decisions, our approach incorporates heterogeneous behavioral incentives. Related work on energy-aware streaming, such as \cite{Turkkan2022GreenABREA}, applies deep-reinforced learning to optimize energy consumption without reducing the user's QoE. In contrast, our model incentivizes the users to pick eco-friendly choices, and the provider leverages their responses to estimate preferences, guiding the system toward progressively greener outcomes.

\section{User Acceptance Model}\label{sec:user acceptance model}

Let $N$ be the set of users, indexed by $n \in \{1, \ldots, N\}$. Each user consumes on-demand digital video and chooses a video resolution between the normal video resolution and the energy-efficient one, which is the green option. We assume that each video has a resolution between a minimum $x_{\min}$ and a maximum $x_{\max}$ bitrates. Let $x_{\ell, n}$, $x_{h, n}$ denote the bitrates of the low and high resolutions for each user, respectively, with $0 \leq x_{\min} \leq x_{\ell, n} < x_{h, n} \leq x_{\max}$. The green action corresponds to selecting $x_{\ell, n}$, and if user $n$ selects it, we call it a green user; it provides a \emph{flexibility} (or energy–consumption reduction) $x_n := x_{h, n} - x_{\ell, n}$. Taking the flexibility for each user, we have $\mathbf{x} = (x_1, x_2, \ldots, x_N)$, the array of flexibilities.

From \cite{Hossfeld2023AGE}, we utilize for each user $n$ a greenness factor $\gamma_n > 1$, as an advantage factor, assuming $n$ rates lower video quality with a higher score if it knows that it saves energy. Formally, a green user is satisfied with maximum bitrate $x'_{\max} = x_{\max} / \gamma_n$, and the QoE model from \cite{Hossfeld2023AGE} is depicted as,
\begin{equation}\label{eq:f_gamma 1}
	f_{\gamma_n}(x) = \frac{4}{\log (x'_{\max}) - \log(x_{\min})}\log (x) + \frac{\log (x'_{\max}) - 5 \log (x_{\min})}{\log (x'_{\max}) - \log (x_{\min})},
\end{equation}
for $x \in [x_{\min}, x_{\max}]$. Rearranging terms, this formula can be written as
\begin{equation}\label{eq:f_gamma 2}
	f_{\gamma}(x) = 1 + 4\left( \frac{\log(x) - \log(x_{\min})}{\log(x_{\max}) - \log \left(\gamma_n x_{\min}\right)}\right).
\end{equation}

\noindent Differentiating \eqref{eq:f_gamma 2} w.r.t. $\gamma_n$ we take $\frac{\partial f_{\gamma_n}(x)}{\partial \gamma_n} > 0 $ for any $\gamma_n > 0$. Hence, a larger $\gamma_n$ indicates an increasing intrinsic valuation of energy efficiency. 

Next, we consider the Mean Opinion Score (MOS) from \cite{Hossfeld2023AGE} as the user's $n$ utility, which we denote with $U_n$, and is normalized in $[0, 1]$. Mathematically, we have $U_n(x) := \frac{1}{5} f_{\gamma_n}(x)$, which is a decreasing function w.r.t. to $\gamma_n$. Then the utility loss when user $n$ moves from high to low resolution is defined as $\Delta U_n := U_n(x_{h, n}) - U_n(x_{\ell, n})$. Substituting the $U_n(x_{h, n})$ and $U_n(x_{\ell, n})$ with $\frac{1}{5} f_{\gamma_n}(x_{h, n})$ and $\frac{1}{5} f_{\gamma_n}(x_{\ell, n})$, respectively, gives

\begin{equation}\label{eq:Delta Un}
	\Delta U_n = \frac{4 \left(\log \left(x_{h, n} \right) - \log \left(x_{\ell, n}\right)\right)}{5 \left(\log \left(x_{\max} \right) - \log \left(\gamma_n x_{\min}\right)\right)}.
\end{equation}

Notably, the model presented in this paper focused on normalized incentives, which can be properly converted to monetary or non-monetary ones by means of a scaling factor.

Further, we introduce the term $s_n$ to be the savings achieved in the energy bill by user $n$ when watching at bitrate $x_{\ell, n}$. The net benefit loss for user $n$, in monetary units, is then $\text{NB}_{loss, n} = \Delta U_n - s_n$. Following \cite{Minou2015IncentivesAT}, if $r_n \geq \text{NB}_{loss, n}$, then the green option is the optimal decision for the user. Therefore, we define the \emph{minimum acceptable incentive} for user $n$ as  $r_{\min, n} = \max \{\text{NB}_{loss, n}, 0\}$. Ideally, the probability $p_n(x_n)$ for user $n$ to take the green option given offer $r_n$ is $p_n(r_n) = 1$ if $r_n \geq r_{\min, n}$, otherwise we have $p_n(r_n) = 0$.

To capture the uncertainty of the user about its own utility function at any given time, we model acceptance as a Bernoulli trial; that is, we assume that acceptance is stochastic. In particular, given an offered incentive $r_n$ to user $n$, the probability that user selects the green option now becomes,
\begin{equation}\label{eq:sigmoid}
    p_n(r_n) = \frac{1}{1 + e^{-\delta_n(r_n - r_{\min, n})}},
\end{equation}
where $p_n(\cdot)$ is a smooth, strictly increasing sigmoid function, satisfying $p_n(0) = 0$, $\lim_{r_n \to \infty} p_n(r_n) = 1$, and $p_n(r_{\min, n}) = \frac{1}{2}$. The slope parameter $\delta_n > 0$ controls responsiveness such that increasing $\delta_n$ produces a behavior increasingly closer to the step-function behavior. Now, the utility loss \eqref{eq:Delta Un} under the stochastic acceptance becomes $p_n(r_n) \cdot \Delta U_n$.

By the construction of the model, the parameter $r_{\min, n}$ is personalized for each user. %, and in fact may also depend on the specific video to be watched as well as on what the green alternative is. 
However, we can consider for simplicity that all videos have the same default resolution as well as the same alternative one, that is $x_{h, n} = x_h$ and $x_{\ell, n} = x_\ell$ for any $n \in N$. In this case, we can take for simplicity that $r_{\min, n}$ is proportional to the duration of the video, with the scaling factor being a personalized parameter. For the value of parameter $\delta_n$, we also consider that it is the same for all videos to be watched by user $n$.

\paragraph*{Provider policy} In this work, the provider offers the incentives $r_n$ and then receives the expected cost and expected total flexibility as
\begin{align}
\label{eq:expected-cost}
	& \mathbb{E}[\mathrm{Cost}] := \sum_{n=1}^N p_n(r_n) \cdot r_n + c_{\mathrm{admin}}\sum_{n=1}^N \mathbf 1\{r_n > 0\},\\
\label{eq:expected-flex}
	& \mathbb{E}[\mathrm{Flexibility}] := \sum_{n=1}^N p_n(r_n) \cdot x_n,
\end{align}
where \(c_{\mathrm{admin}} \ge 0\) is a per-offer administrative cost.

Finally, the provider's optimal policy is the allocation of $r_n$s that maximizes the ratio between the expected total flexibility and the expected cost (in short, Flexibility-Cost ratio).

\section{Extensions}\label{sec:extensions}

\subsection{Altruistic Behavior}\label{subsec:altruistic behavior}

Here, we shift our attention to the case where some users internalize the collective environmental benefit. Let $\beta_n \in [0, 1]$ denote the weight user $n$ places on its individual utility. First, we define the social green-wellbeing as the mean of the utilities, that is $SW(\mathbf{x}) := \frac{1}{N}\sum^N_{n=1} U_n(x)$. The user $n$ experiences the composite utility $\widetilde{U}_n(x) := \beta_n U_n(x) + (1 - \beta_n)SW(\mathbf{x})$. Then, the minimum acceptance incentives for the altruistic user become, $r^{\beta_n}_{\min, n} = \widetilde{U}_n(x_{h, n}) - \widetilde{U}_n(x_{\ell, n})$.

Notably, $\gamma_n$ has an implicit effect on the altruistic behavior. In particular, higher $\gamma_n$ also steepens the MOS curve and increases $\Delta U_n$ for a fixed pair of $(x_{\ell, n}, x_{h, n})$, see equation \eqref{eq:Delta Un}. On the other hand, from equation \eqref{eq:f_gamma 2}, if $\gamma_n$ increases then $U_n(x)$ increases, but is clipped in $[0, 1]$. Due to opposite signs of $U_n$ and $SW(\mathbf{x})$ in $\widetilde{U}(x)$, the overall effect on required incentives can be non-trivial. In more detail, higher $\gamma_n$ values both increase the perceived private QoE loss, pushing upwards the $r_{\min, n}$ and may decrease the minus signed social well-being term, pushing downwards $r_{\min, n}$ if altruism is present. We continue by examining two cases, i) heterogeneous case (all users are different), and ii) the two-groups population case (users partitioned into groups with different characteristics).

\paragraph{Heterogeneous population} Assume that $U_n(x) = U_m(x)$ for any $n, m \in N$, with $n \neq m$. The minimum acceptable incentive for user $n$ under altruism $\beta_n$ is $r^{\beta_n}_{\min, n} = \beta_n \Delta {U}_n(x) + \frac{1 - \beta_n}{N} \sum^{N}_{i=1}\Delta {U}_i(x)$. Since $\Delta U_i(x) \geq 0$ for any $i \in N$, and the linear relationship between $r^{\beta_n}_{\min, n}$ and $\beta_n$, minimizing $r^{\beta_n}_{\min, n}$ w.r.t. $\beta_n$ is attained when $\beta_n = 1$ if $ \Delta {U}_n(x) < \frac{1}{N} \sum^{N}_{i=1}\Delta {U}_i(x)$. Otherwise, we have $\Delta {U}_n(x) > \frac{1}{N}\sum^{N}_{i=1}\Delta {U}_i(x)$, and $r^{\beta_n}_{\min, n}$ is minimized for $\beta_n = 0$. In other words, when the other users provide more flexibility on average than $n$, $r^{\beta_n}_{\min, n}$ is minimized if the user $n$ becomes selfish. When the user $n$ provides more flexibility than the average of all users, then $r^{\beta_n}_{\min, n}$ is minimized if the user $n$ becomes fully altruistic.

\paragraph{Two-groups population} Assume that the set of users $N$ is partitioned into two subsets $L$ and $M$, and the users at each subset have the same utility but different $\gamma_n$, that is $U^L_i(x) = U^L_j(x)$ for any $i,j \in L$, and $U^M_i(x_i) = U^M_j(x)$ for any $i,j \in M$. Abusing notation, we denote as $SW^L(\mathbf{x})$, and $SW^L(\mathbf{x})$ the case where we consider the social well-being within groups $L$ and $M$, respectively. The social green-wellbeing now becomes $SW(\mathbf{x}) := (L \cdot SW^L(\mathbf{x}) + M \cdot SW^M(\mathbf{x}))/N$. Simple algebra gives 
\[
r^{\beta_n}_{\min, n} = \Biggl\{
\begin{array}{ll}
\left(\beta_n + \frac{(1 - \beta_n) L}{N}\right) \Delta U^L(x) + \frac{(1 - \beta_n) M}{N} \Delta U^M(x), & n \in L,  \\
\frac{(1 - \beta_n) L}{N} \Delta U^L(x) + \left(\beta_n + \frac{(1 - \beta_n) M}{N}\right) \Delta U^M(x), & n \in M.
\end{array}
\]

\subsection{Educating users for reductions of $r_{\min,n}$}\label{subsec:education-rmin}

Here, we model education as a driver of greener user choices. Exposure to information campaigns or repeated feedback can raise a user’s intrinsic satisfaction with eco-friendly actions, effectively increasing $s_n$ as their perceived non-monetary benefit grows. Learning effects documented in behavioral-economics experiments \cite{Erev2014MaximizationLA} likewise show that sustained environmental messaging can reduce energy use independently of financial motives, which can be modeled as an increase in $\gamma_n$.

If the provider targets a minimum acceptance incentive $r^{\text{target}}_{\min, n}$ per-user, then for each user $n$ it must hold $r^{\text{target}}_{\min, n} = \Delta U_n - s_n$. Solving for $x_{\ell,n}$, $\log(x_{\ell,n}) = \log(x_{h, n}) - \frac{5}{4}(\log(x_{\max}) - \log(\gamma_n x_{\min})) \cdot \Delta U_n$. Exponentiating and substituting $\Delta U_n$ the expression boils down to
\begin{equation}\label{eq:educate}
	x_{\ell} = x_{h} \cdot \exp\left(-\left(\frac{5}{4}(\log(x_{\max}) - \log(\gamma_n \cdot x_{\min})\right) \cdot ({r^{\text{target}}_{\min, n} + s_n}))\right).
\end{equation}
From equation \eqref{eq:educate} we observe that the lower resolution decreases as ${r^{\text{target}}_{\min, n}}$, $\gamma_n$ and $s_n$ increase. On the other hand, lowering $r_{\min, n}$ can be achieved either by increasing $s_n$, from $r^{\text{target}}_{\min, n} = \Delta U_n - s_n$, or by increasing $\gamma_n$ from \eqref{eq:Delta Un}. 

Another way to reduce $r_{\min, n}$ for a population of users is to require a milder quality drop. Because, for each user $n$, $\Delta U_n$ is a strictly increasing function w.r.t. to $\log \left( \frac{x_h}{x_{\ell}} \right)$. Therefore, to achieve the required incentivization, the provider can either invest resources to lower users' $r_{\min, n}$ values, or request a smaller bitrate reduction. Hence, policymakers can set outreach goals—such as achieving a specific rise in measured green-awareness scores—to meet a measurable Flexibility-Cost ratio in incentive budgets.

\subsection{Learning from data}\label{subsec:learning from data}

Now, consider the setting where for each user $n$ the provider observes $m$ past interactions $\mathcal{D}_m = \{(r_{n, i}, \chi_{n,i}) \mid i \in [m]\}$, where the offered incentives $r_{n, i}$ are sampled from the known distribution $\nu$, and the response $\chi_{n,i} \in \{0, 1\}$ is drawn as $P[\chi_{n, i} = 1 \mid r_{n, i}] = p_n(r_{n, i})$. The samples are i.i.d. conditional on the true parameters. The provider wants to estimate the true minimum acceptance incentives $r_{\min, n}$ for each user. For that, we employ logistic regression to estimate each user's $r_{\min, n}$ during the learning state. Formally we have $P[\chi = 1 \mid r; \theta_n] = p_n(\theta_{n, 0} + \theta_{n,1}r)$, and $p_n(\cdot)$ follows the equation $\eqref{eq:sigmoid}$. For the fitting, we maximize the likelihood
\[
    \widehat{\theta}_n = \argmax_{\theta \in \mathbb{R}^2} \sum_{i \in m} \chi_{n, i} \log(z_{n, i}) + (1 - \chi_{n, i}) \log(1 - z_{n, i}),
\]
where $z_{n, i} = p_n(\theta_{n, 0} + \theta_{n,1}r)$, $\theta_{n, 1} = \delta_n$, and $r_{\min, n} = - \frac{\theta_{n, 0}}{\theta_{n, 1}}$. Thus, the maximum-likelihood estimate $\widehat{\delta}_n = \widehat{\theta}_{n, 1}$, and $\widehat{r}_{\min, n} = - \frac{\widehat{\theta}_{n, 0}}{\widehat{\theta}_{n, 1}}$. Under standard regularity conditions, the maximum-likelihood estimator is strongly consistent, see \cite{Casella2024StatisticalI}, implying that as $m$ increases the estimation errors for $\widehat{r}_{\min, n}$ and $\widehat{\delta}_n$ decreases, see Figures \ref{fig:Mean absolute error uniform} and \ref{fig:Mean absolute error normal}. Further, identifiability and low estimation variance require that the support of $\nu$ covers a neighborhood of the true $r_{\min, n}$; otherwise, the model extrapolates and estimates become unstable. Regularization or hierarchical pooling across similar users can reduce variance in small-sample regimes.

\section{Evaluation}\label{sec:evaluation}

\subsection{Synthetic data generator.}\label{subse:synthetic data generator}
To evaluate the user acceptance model, we designed a synthetic data generator that produces a population of $N = 1000$ users. Each user's attributes mirror real-world variations in video-streaming behavior, environmental awareness, and incentive responsiveness. At first, \emph{high/low bitrates} $(x_{h,i},x_{l,i})$ are drawn from discrete sets $\{2000,3000,4000,5000\}\;\mathrm{kbps}$ and $\{300,600,1200,1500\}\;\mathrm{kbps}$, respectively. Offered incentives are sampled under two regimes: i) a uniform distribution $\mathcal{U}[a, b]$, and ii) a normal distribution $\mathcal{N}(\mu, \sigma^2)$. Within each regime, we vary the parameters $\beta_n$, $\gamma_n$, $\delta_n$ as specified in the individual experiments. Furthermore, $c_{\text{admin}} = 0.04$ and, following \cite{Hossfeld2023AGE}, the $U_n$ values are clipped in $[0, 1]$. In each plot, the dashed lines correspond to an optimal provider's policy. 

\subsection{Optimal incentives.}
\begin{figure}
    \centering
    $\begin{array}{cc}
     \includegraphics[scale=.35]{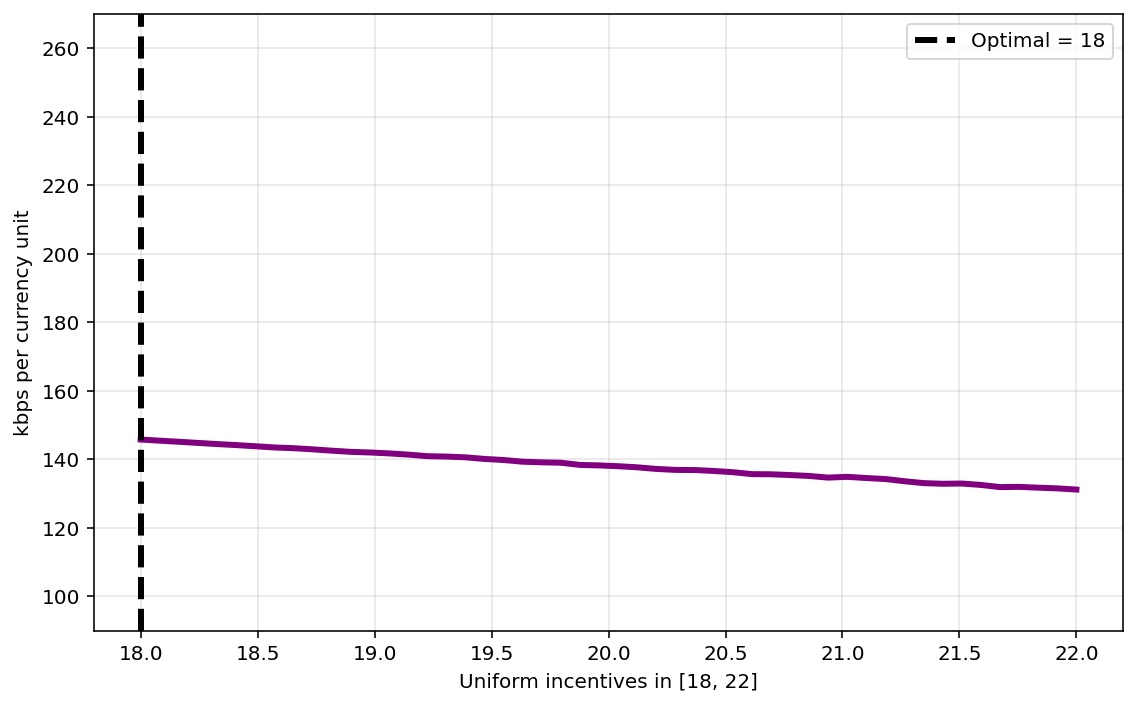} & \includegraphics[scale=.35]{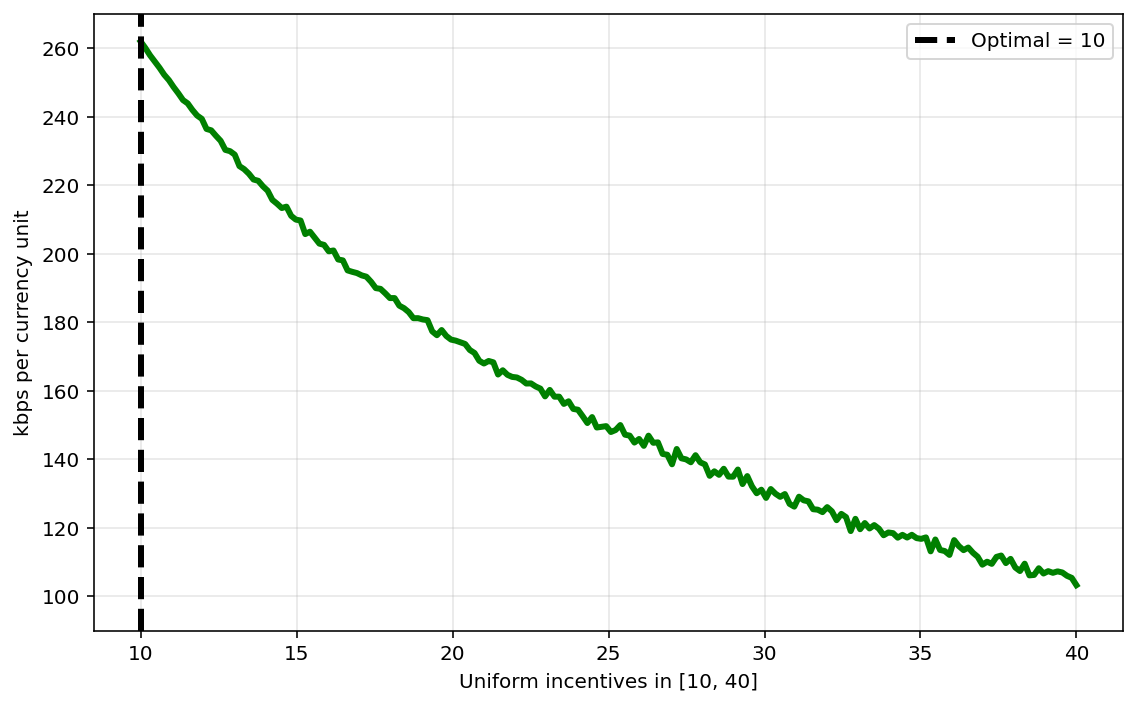}
    \end{array}$
    \caption{Flexibility-Cost ratio vs the discreptives of the offered incentives $\sim \mathcal{U}[a, b]$. $a = 18$, $b = 22$ (left). $a = 10$, $b = 40$ (right).} 
    \label{fig:incentivized users uniform descriptives}
\end{figure}

\begin{figure}
    \centering
    $\begin{array}{cc}
     \includegraphics[scale=.35]{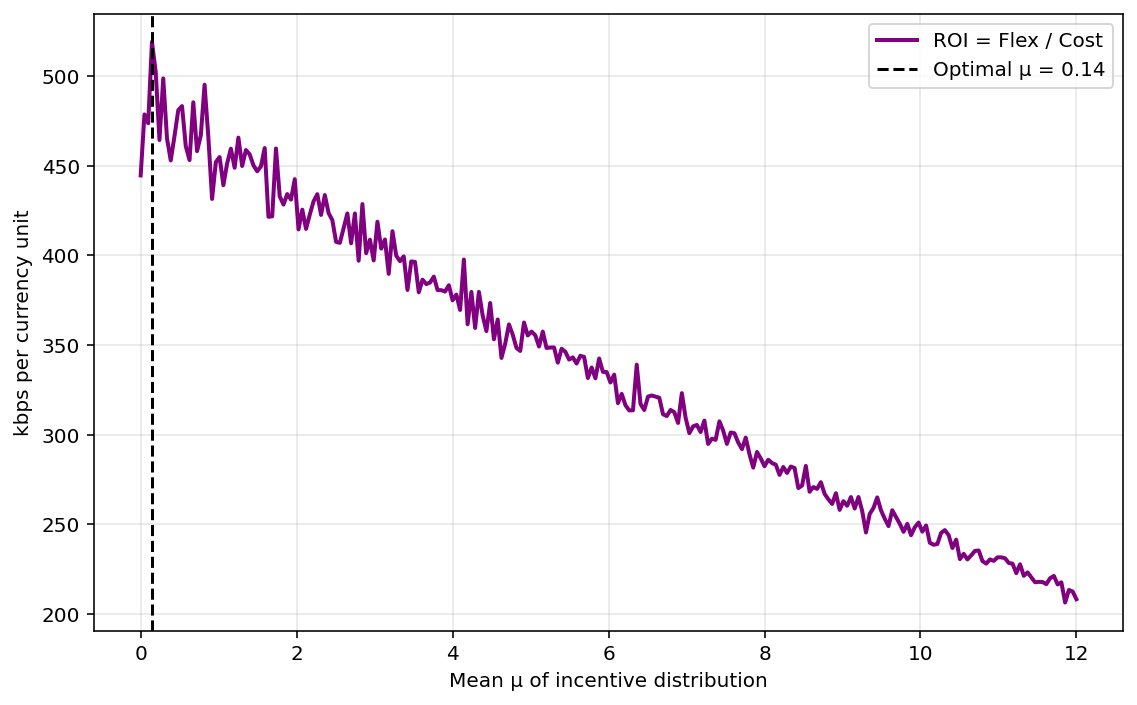} & \includegraphics[scale=.35]{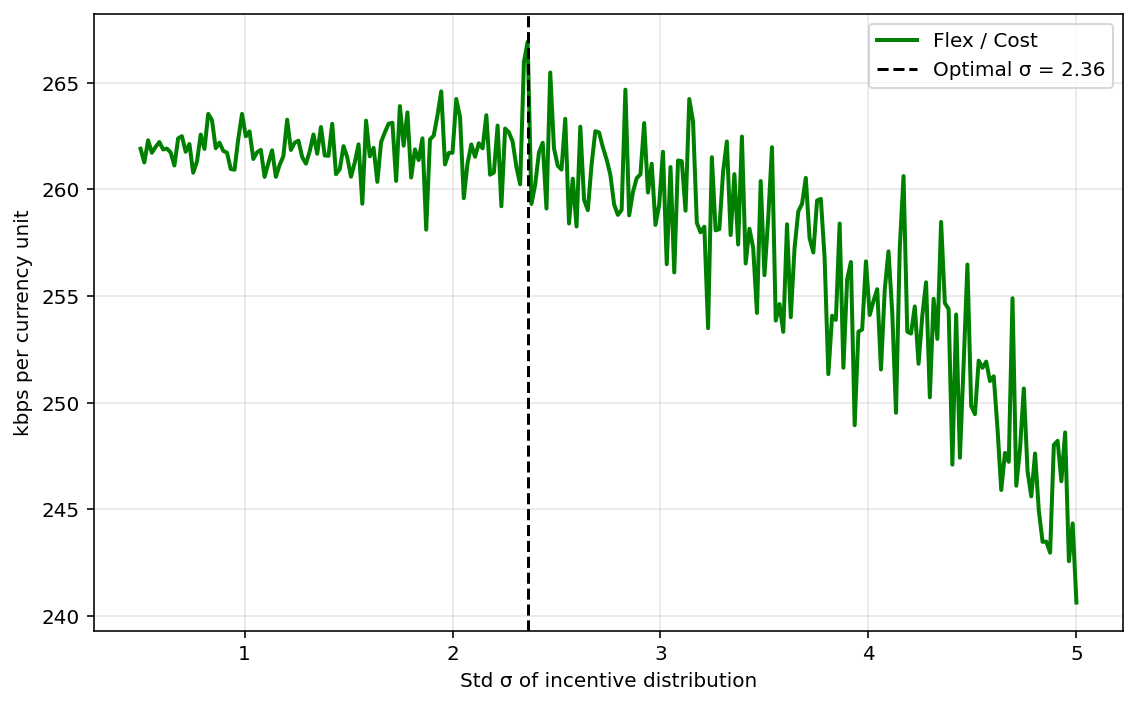}
    \end{array}$
    \caption{Flexibility-Cost ratio vs the descriptives of the offered-incentives $\sim \mathcal{N}(\mu, \sigma^2)$. $\mu \in [0, 12]$ and $\sigma = 5$ (left). $\mu = 10$ and $\sigma \in [0.5, 5]$ (right).} 
    \label{:incentivized users normal descriptives}
\end{figure}

We first study the case where incentives are offered to all users, sampled either from the uniform or the normal distribution above. For each distribution, we determine the optimal offered incentive level, defined as the value of the offered incentive that maximizes the Flexibility-Cost ratio. Throughout this analysis, we fix $\delta_n$ to be the same for each $n \in N$, $\gamma_n = 1.15$, and $\beta_n = 1$. In Figures \ref{fig:incentivized users uniform descriptives} and \ref{:incentivized users normal descriptives}, the purple and green curves depict the Flexibility-Cost ratio. Figure \ref{fig:incentivized users uniform descriptives} compares two ranges of $\mathcal{U}[a, b]$: $a=18$, $b=22$ (left), and $a=10$, $b=40$ (right). We observe that the Flexibility-Cost ratio deteriorates when we shift to larger offered incentives. In both cases, the provider's optimal policy is to offer as few incentives as possible.

In Figure \ref{:incentivized users normal descriptives} we have $\mathcal{N}(\mu, \sigma^2)$: $\mu \in [0, 12]$ and $\sigma = 5$ is fixed (left), and $\mu = 10$ fixed and $\sigma \in [0.5, 5]$ (right). Contrary to the uniform case, here the provider's optimal policy depends on the deviations of the offered incentives. We observe that, keeping $\sigma$ fixed, the optimal policy of the provider is when $\mu$ has small values. On the other hand, when the pool of users has fixed $\mu$, the deviation that optimizes the provider's policy is when $\sigma$ has intermediate values.

\subsection{Number of incentivized users.}\label{subsec:number of incentivized users}
\begin{figure}
    \centering
    $\begin{array}{cc}
     \includegraphics[scale=.35]{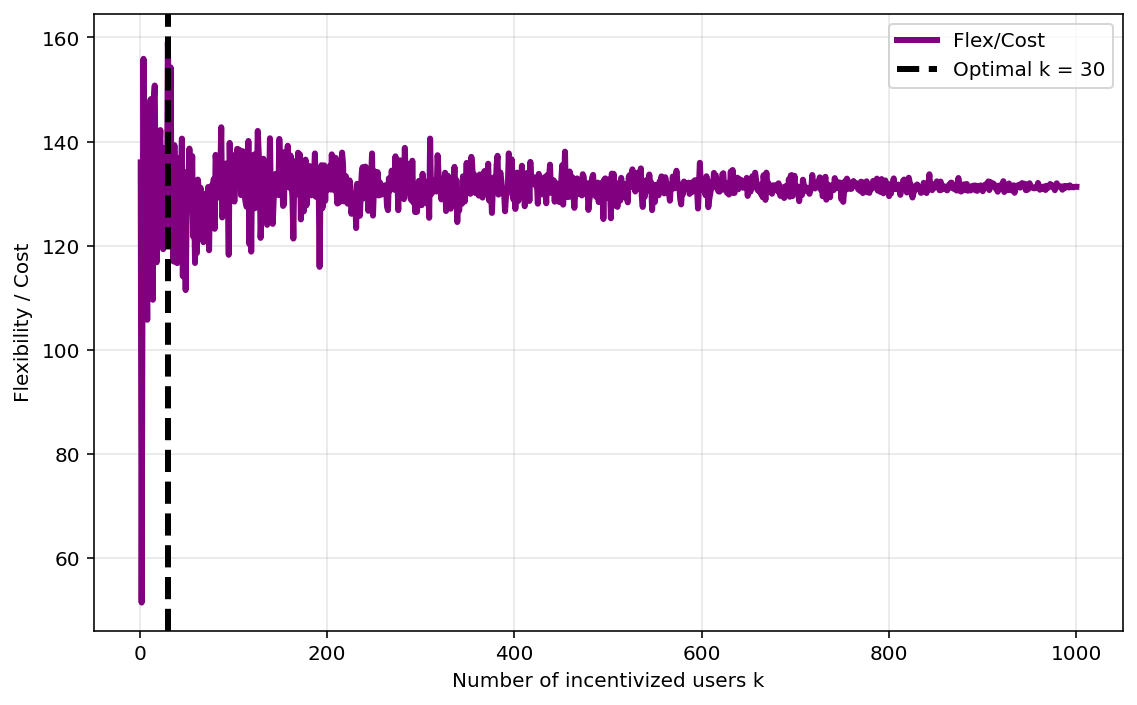} & \includegraphics[scale=.35]{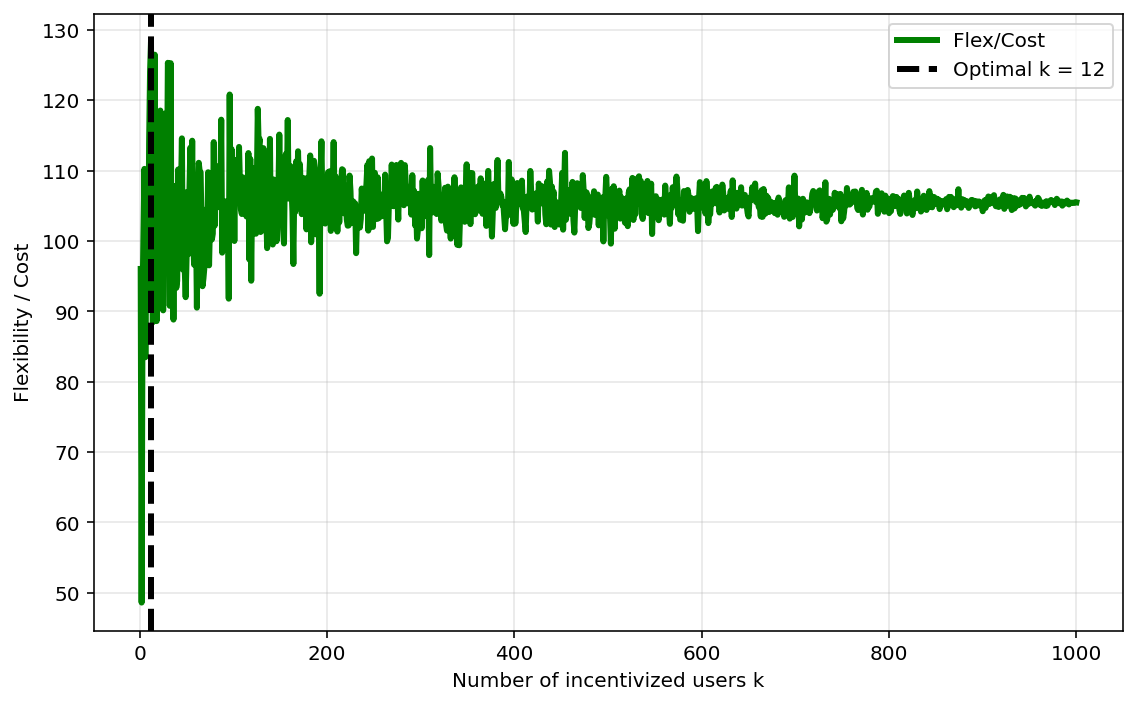}
    \end{array}$
    \caption{Flexibility-Cost ratio vs the numbers of incentivized users. The offered-incentives sampling from the uniform distribution, with $\mathcal{U}[18, 22]$ (left), and $\mathcal{U}[10, 40]$ (right).} 
    \label{fig:incentivized users uniform}
\end{figure}

We now examine how the number of incentivized users influences the provider’s optimal policy, for fixed $\delta_n = 1.2$. Figure \ref{fig:incentivized users uniform}, $r_n \sim \mathcal{U}[a,b]$, shows that as the number of incentivized users increases, the expected total flexibility and the expected flexibility have the same behavior. For small user counts, the Flexibility-Cost ratio fluctuates, but the oscillations narrow and stabilize as the population expands, independently of the range in the uniform distribution. A second remark is that wider incentive ranges (e.g, $b - a$) yield larger oscillations and shift the provider’s optimum toward fewer incentivized users. 

\begin{figure}
    \centering
    $\begin{array}{ccc}
     \includegraphics[width=0.3\linewidth]{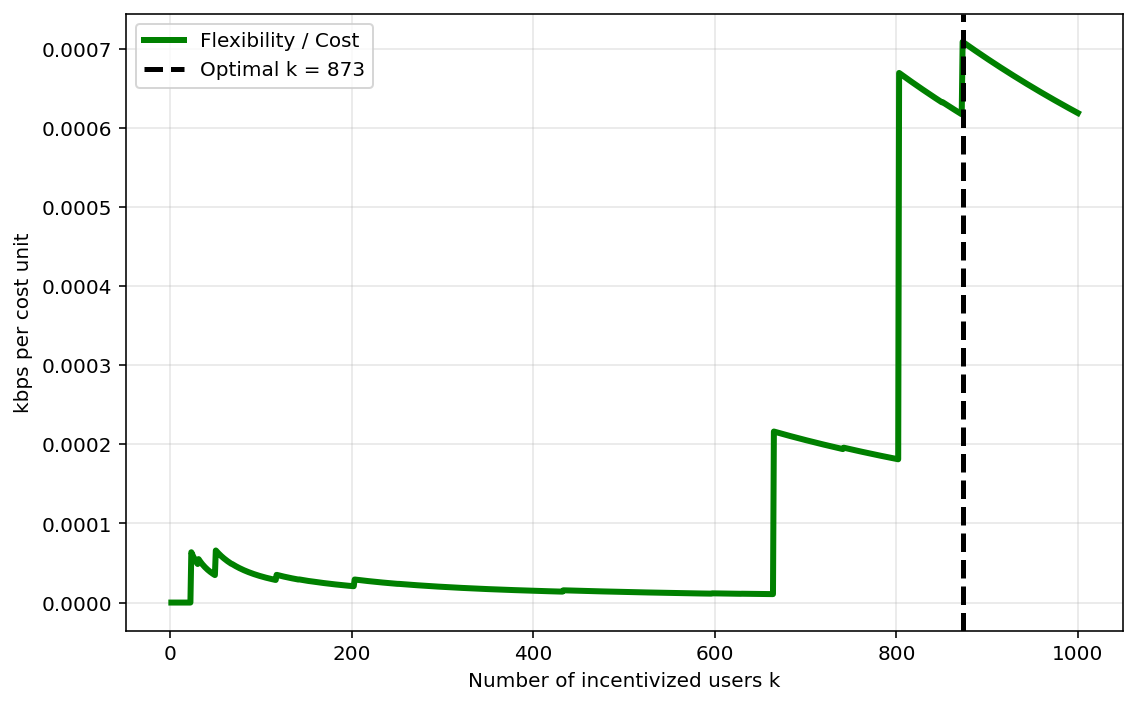} & \includegraphics[width=0.3\linewidth]{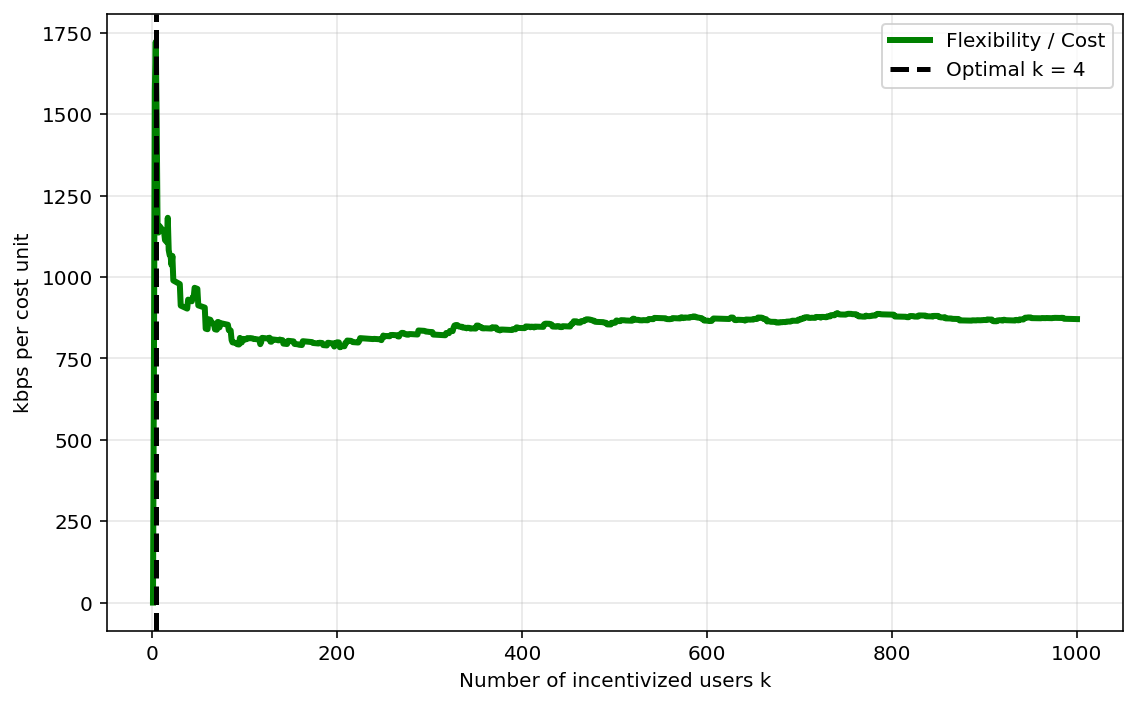} & \includegraphics[width=0.3\linewidth]{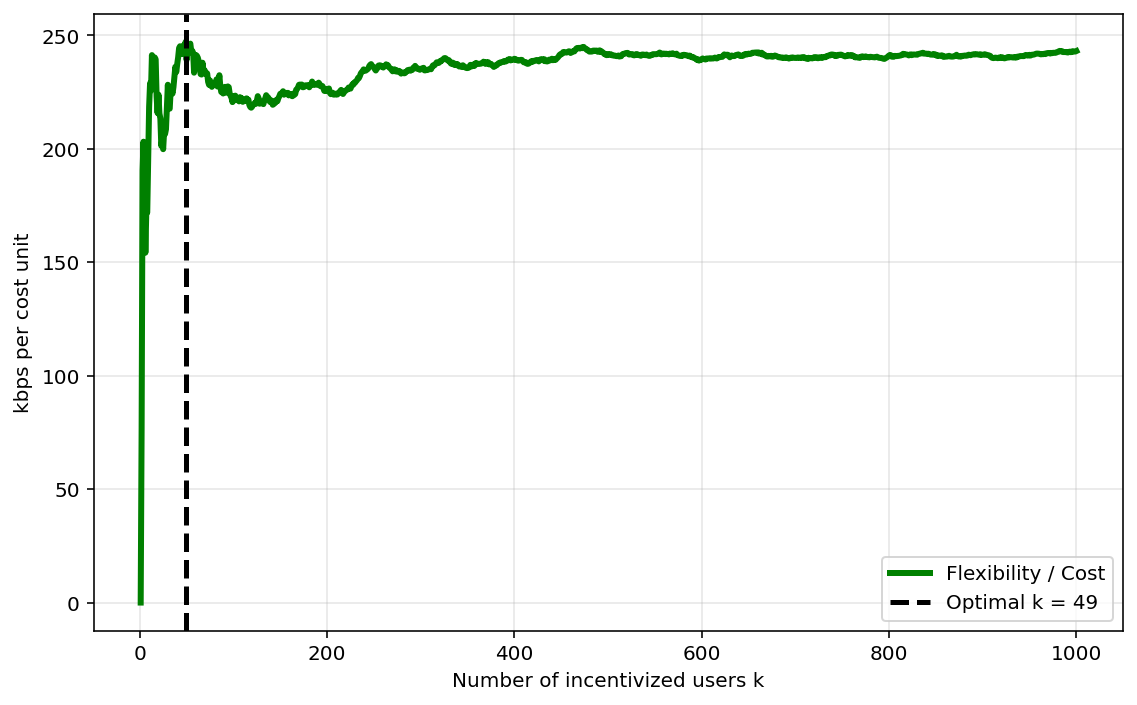} \\
    \end{array}$
    \caption{Flexibility-Cost ratio vs number of incentivized users. Offered-incentives are sampled from the normal distribution, with  $\mathcal{N}(0.1, 0.05)$ (left), $\mathcal{N}(1, 1)$ (left), and  $\mathcal{N}(10, 1)$ (right).} 
    \label{fig:incentivized users normal}
\end{figure}

In Figure \ref{fig:incentivized users normal}, $r_n \sim \mathcal{N}(\mu, \sigma^2)$, we observe that with low incentives ($\mu = 0.1$, left), the provider’s best policy is to incentivize nearly all users ($\approx 873$). Increasing the mean ($\mu = 1$, middle) or variance $\sigma$ moves the optimum to progressively smaller subsets.

\begin{figure}
    \centering
    $\begin{array}{cc}
     \includegraphics[scale=0.45]{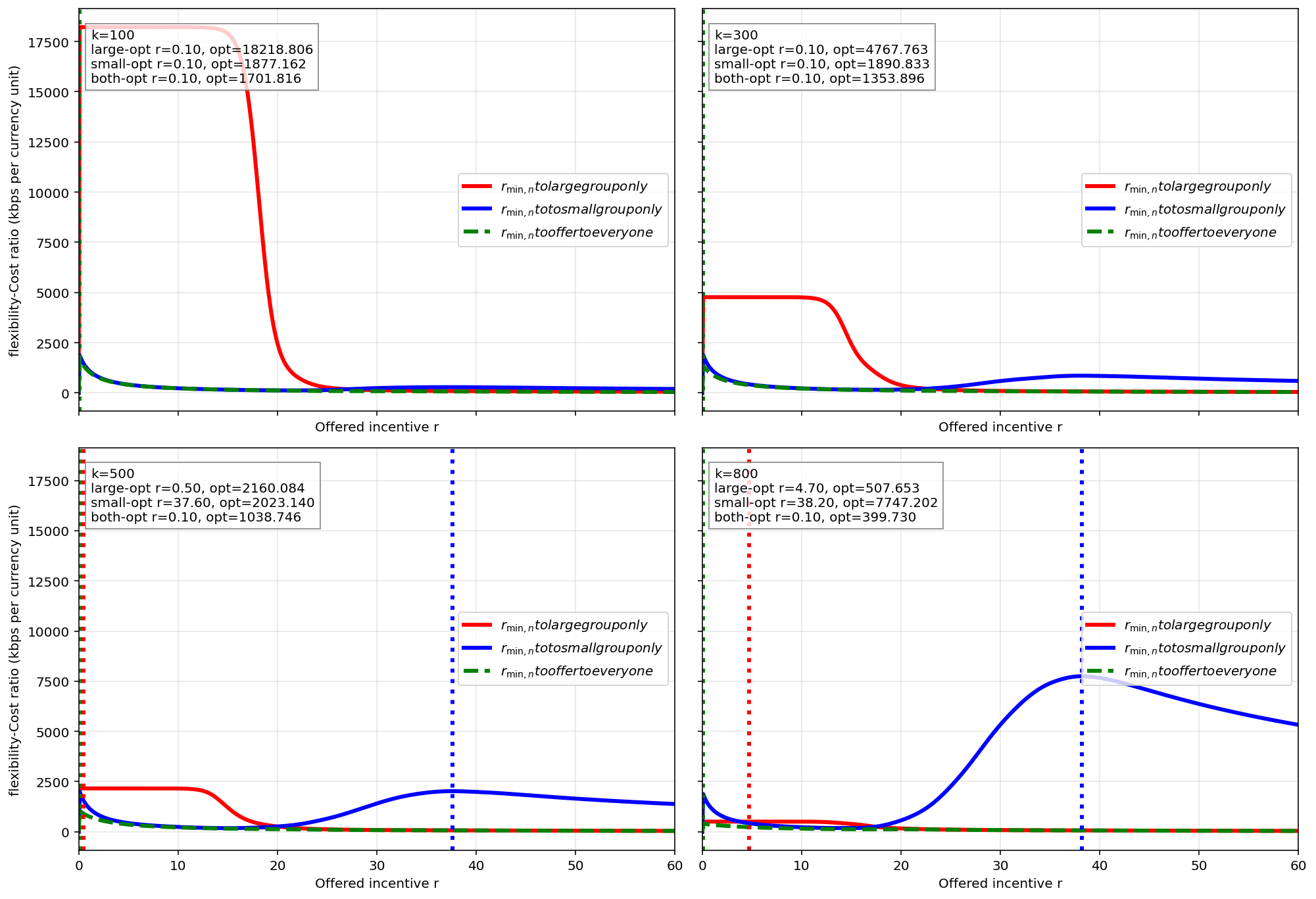}
    \end{array}$
    \caption{Provider offers $\mathcal{N}(\mu, \sigma^2)$: i) group with small $r_{\min, n}$ (blue), ii) group with large $r_{\min, n}$ (red), and iii) mean $r_{\min, n}$ between groups (green). (top-left) $k = 100$, (top-right) $k = 300$, (bottom-left) $k = 500$, and (bottom-right) $k = 800$.}
    \label{fig:large vs small incentive group}
\end{figure}

Further, we partition $N$ into two groups, one of size $k$, where users have $r_{\min, n} \sim \mathcal{N}(30, 25)$, and the rest where users have $r_{\min, n} \sim \mathcal{N}(3, 0.25)$. Figure \ref{fig:large vs small incentive group} shows that higher values in the curve are better for the provider as it aims to save more kbps per unit cost. Clearly, for the group with a larger $r_{\min, n}$, the curve picks at a higher offered $r_n$, compared to the group with a smaller $r_{\min, n}$. The optimal policy depends strongly on group size $k$. When the large group is small ($k=100$, top-left panel in Figure \ref{fig:large vs small incentive group}), it is better to incentivize the small group. Contrarily, when $k$ is large, the provider might prefer targeting the small group, or offer to everyone at intermediate $r_n$ values.

\begin{figure}
    \centering
    $\begin{array}{cc}
     \includegraphics[scale=.45]{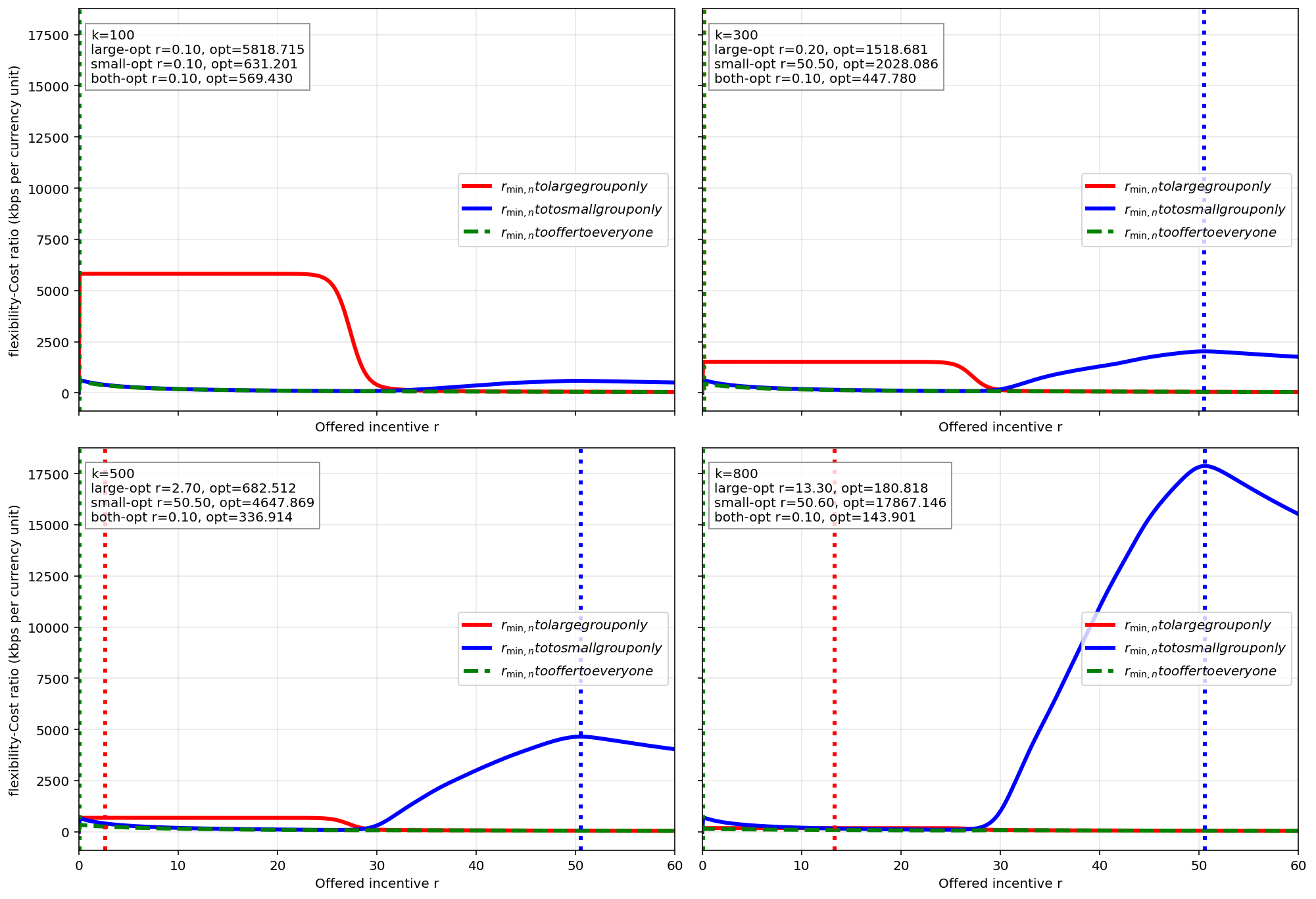}
    \end{array}$
    \caption{Provider offers $\mathcal{U}[a,b]$: i) group with small $r_{\min, n}$ (blue), ii) group with large $r_{\min, n}$ (red), and iii) mean $r_{\min, n}$ between groups (green). (top-left) $k = 100$, (top-right) $k = 300$, (bottom-left) $k = 500$, and (bottom-right) $k = 800$.}
    \label{fig:large vs small incentive group uniform}
\end{figure}

\subsection{Mean vs individual incentives.}\label{subsec:mean vs individual incentives}
\begin{figure}[t]
    \centering
    $\begin{array}{cc}
     \includegraphics[scale=0.35]{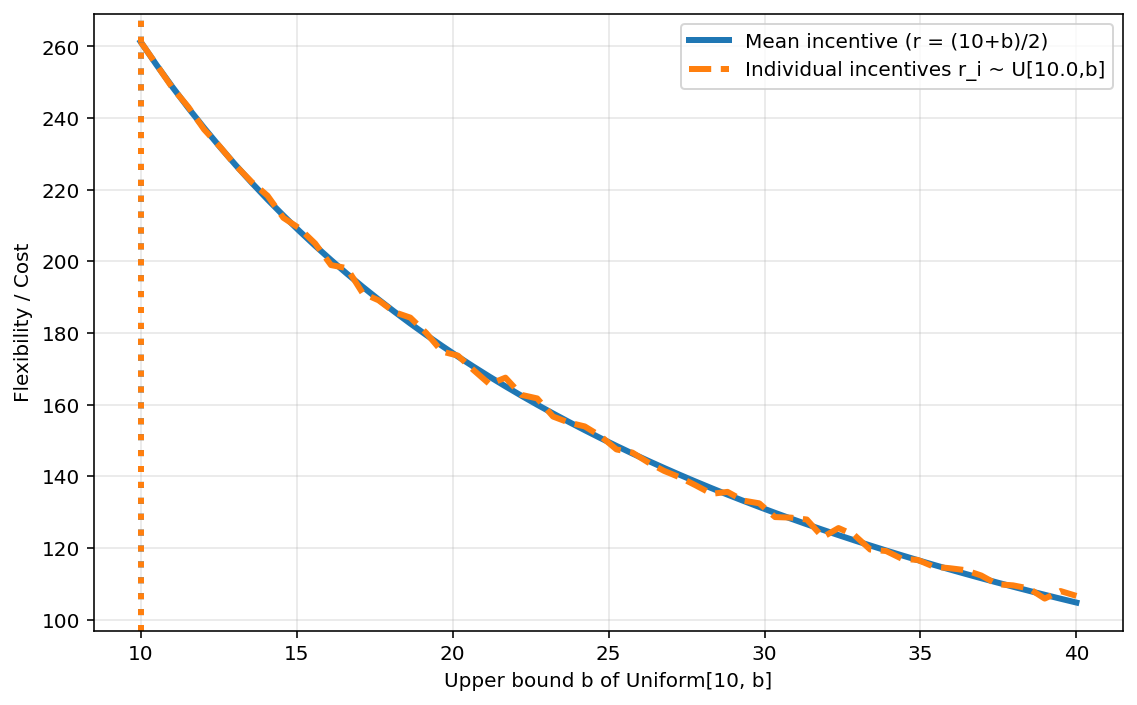} & \includegraphics[scale=0.35]{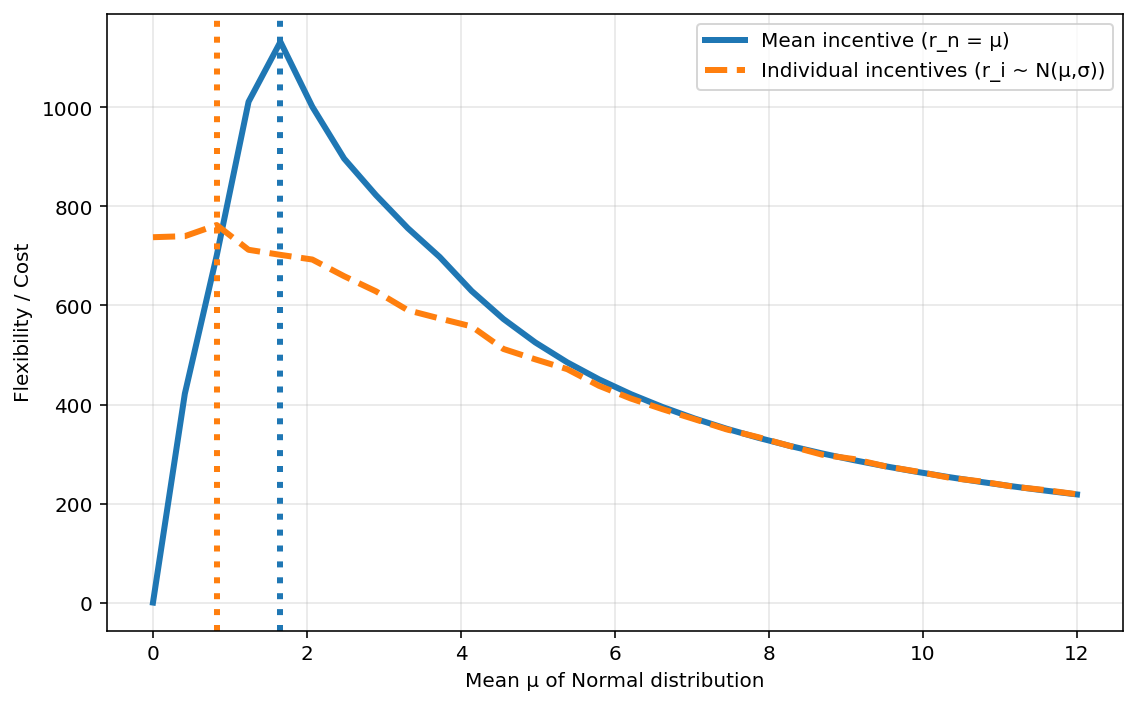}
    \end{array}$
    \caption{Provider offers: mean incentives to all users (orange), individual incentives (blue), sampled from $\mathcal{U}[10, b]$, with $b \in [10, 40]$ (left), and the $\mathcal{N}(\mu, \sigma^2)$, with $\mu \in (0, 12]$ and $\sigma = 2$ (right).}
    \label{fig:mean vs individual}
\end{figure}

\begin{figure}[t]
    \centering
    $\begin{array}{cc}
     \includegraphics[scale=0.35]{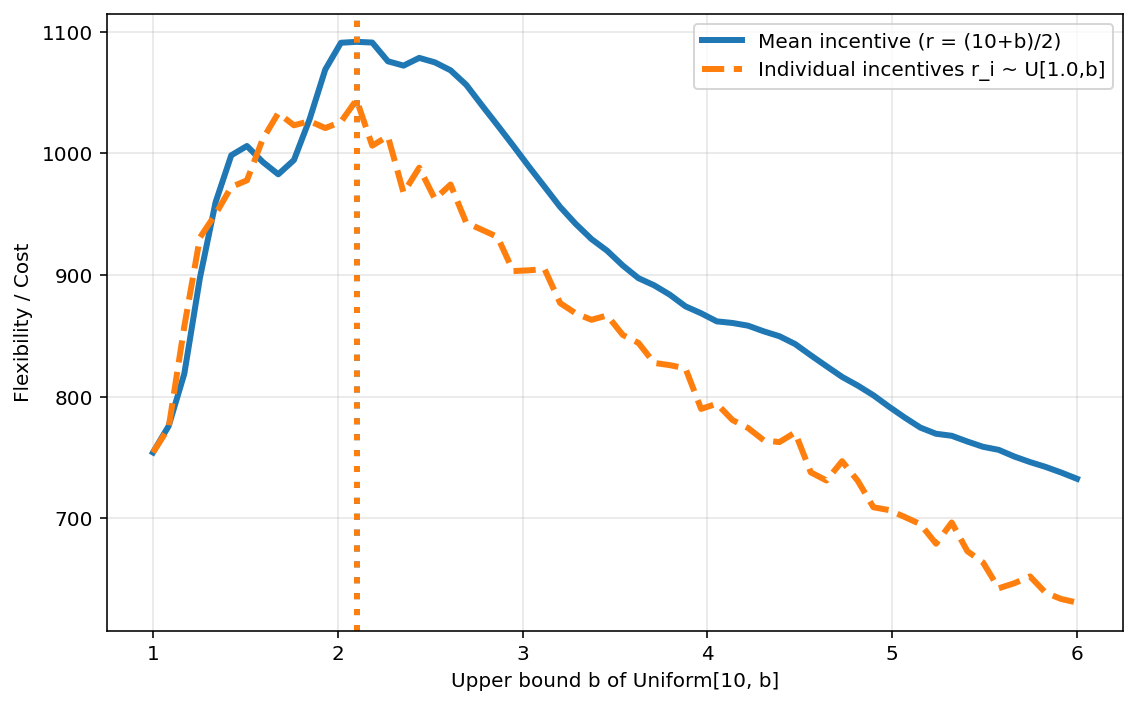} & \includegraphics[scale=0.35]{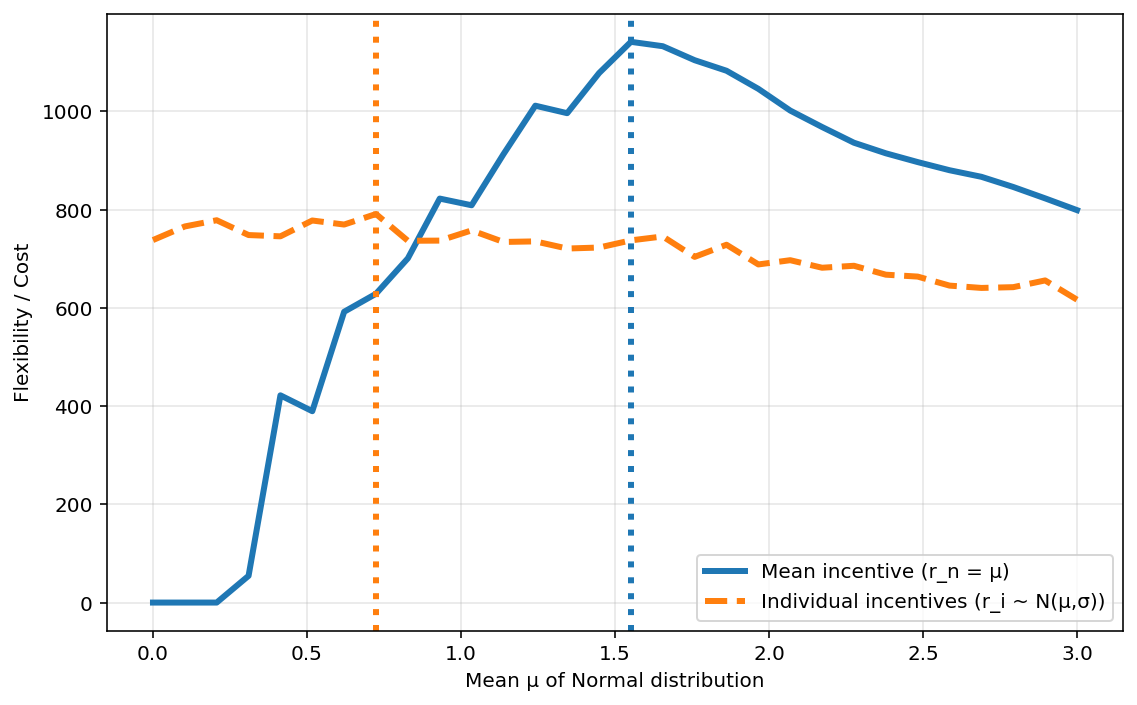}
    \end{array}$
    \caption{Provider offers: mean incentives to all users (orange), individual incentives (blue), sampled from $\mathcal{U}[1, b]$, with $b \in [1, 6]$ (left), and the $\mathcal{N}(\mu, \sigma^2)$, with $\mu \in (0, 3]$ and $\sigma = 2$ (right).}
    \label{fig:mean vs individual 2}
\end{figure}

Here we compare the ``mean-offered'' incentives policy, that is, the provider offers incentives equal to the mean of a distribution to all users, with individualized offers drawn from the same distribution. For that, we take $\delta_n = 100$. In Figure \ref{fig:mean vs individual}, $r_n \sim \mathcal{U}[a, b]$, the provider samples $r_n$ from $\mathcal{U}[10, b]$, with $b \in [10, 40]$. The two policies perform nearly identically, and the provider minimizes $\mathbb{E}[\textrm{Cost}]$ by offering the lowest feasible incentive. In Figure \ref{fig:mean vs individual 2}, the provider samples $r_n$ from $\mathcal{U}[1, b]$, with $b \in [1, 6]$, and attains its optimal incentivization form $b \approx 2.1$. For $b < 2.1$, not many users accept the incentives, while for $b > 2.1$ the $\mathbb{E}[\textrm{Cost}]$ increases, see Figure \ref{fig:mean vs individual 2}. Again, the Flexibility-Cost ratio for the mean-offered incentives policy has the same behaviour as the individual-offered incentives policy. In Figure \ref{fig:mean vs individual}, $r_n \sim \mathcal{N}(\mu, \sigma^2)$, the provider samples $r_n$ from $\mathcal{N}(\mu, \sigma^2)$, with $\mu \in [0, 12]$ and $\sigma = 2$. The Flexibility-Cost ratio initially increases as higher $\mu$ values imply higher $r_n$ and greater $\mathbb{E}[\textrm{Flexibility}]$. Then it is maximized at $\mu = 1.8$, and then decays as $\mathbb{E}[\textrm{Cost}]$ increases. We show similar behavior for the case where $r_n \sim \mathcal{N}(\mu, \sigma^2)$, with $\mu \in [0, 3]$ and $\sigma = 2$. Contrarily, when the provider offers the mean $r_n$, then there are no significant changes in the Flexibility-Cost ratio, Figures \ref{fig:mean vs individual} and \ref{fig:mean vs individual 2}. 

\subsection{Altruistic behavior.}
\begin{figure}[t]
    \centering
    $\begin{array}{cc}
     \includegraphics[scale=0.35]{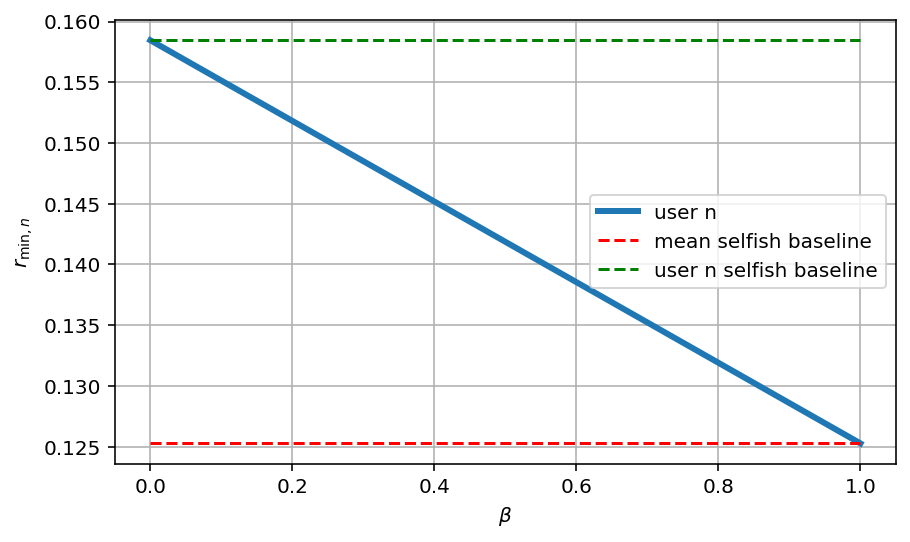} & \includegraphics[scale=0.35]{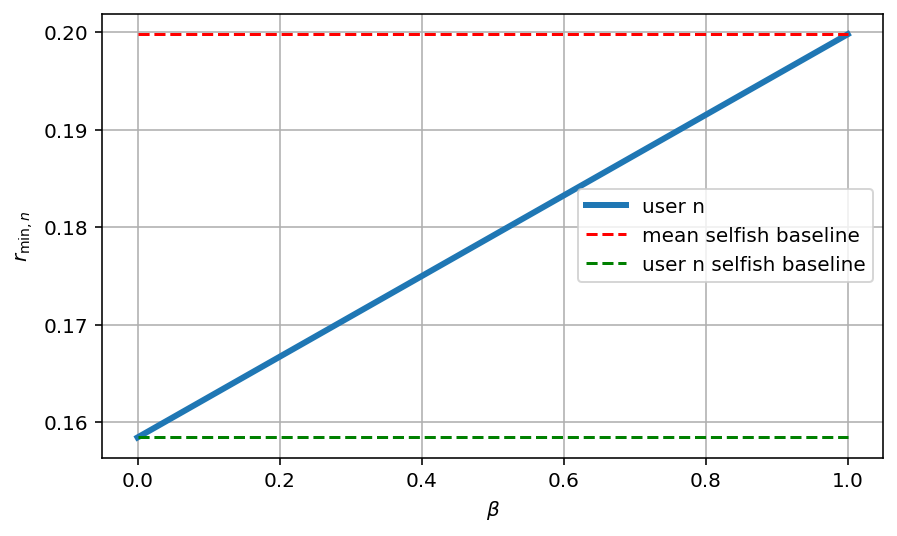}
    \end{array}$
    \caption{User with higher selfish $r_{\min,n}$ compared to the mean $r_{\min,n}$ (left). User with lower selfish $r_{\min,n}$ compared to the mean $r_{\min,n}$ (right), where $r_{\min,n}$ is the red dashed line.}
    \label{fig:altruism}
\end{figure}

\begin{figure}[t]
    \centering
     \includegraphics[scale=0.45]{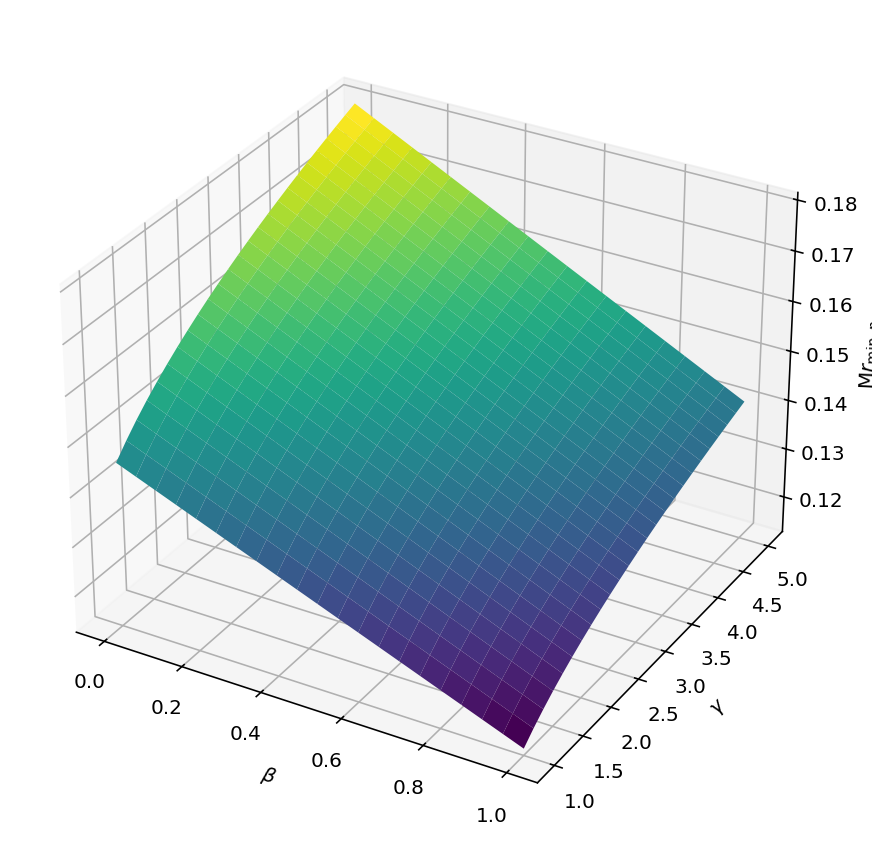} 
    \caption{Relationship between $\beta_n$, $\gamma_n$ and the average $r_{\min,n}$.}
    \label{fig:altruism b g}
\end{figure}

Here we evaluate the effect of the altruism level $\beta_n$ on the $r^{\beta_n}_{\min, n}$ for the user $n$. Figure \ref{fig:altruism} shows that when $n$ has a greater $r_{\min, n}$ than the average, then social concerns deteriorate the $r_{\min, n}$ values, leading to more green actions. Contrarily, when $n$ has a smaller $r_{\min, n}$ than the average, then social concerns increase the $r_{\min, n}$ values. Therefore, to implement altruism optimally, we need to consider the relative positions of the $r_{\min, n}$ compared to the average minimum acceptance incentives. The effect of $\beta_n$ and $\gamma_n$ on the $r_{\min, n}$ is presented in Figure \ref{fig:altruism b g}. Here we sample $x_{\ell,n}$ and $x_{h,n}$, $\beta_n, \gamma_n$ are equal for all users and we compute the $r^{\beta_n}_{\min, n}$. We observe that independent from the $\gamma_n$, if $\beta_n$ increases $r^{\beta_n}_{\min, n}$ decreases, leading to greener actions.

\subsection{Educate user's behavior}
\begin{figure}
    \centering
    $\begin{array}{cc}
     \includegraphics[scale=0.35]{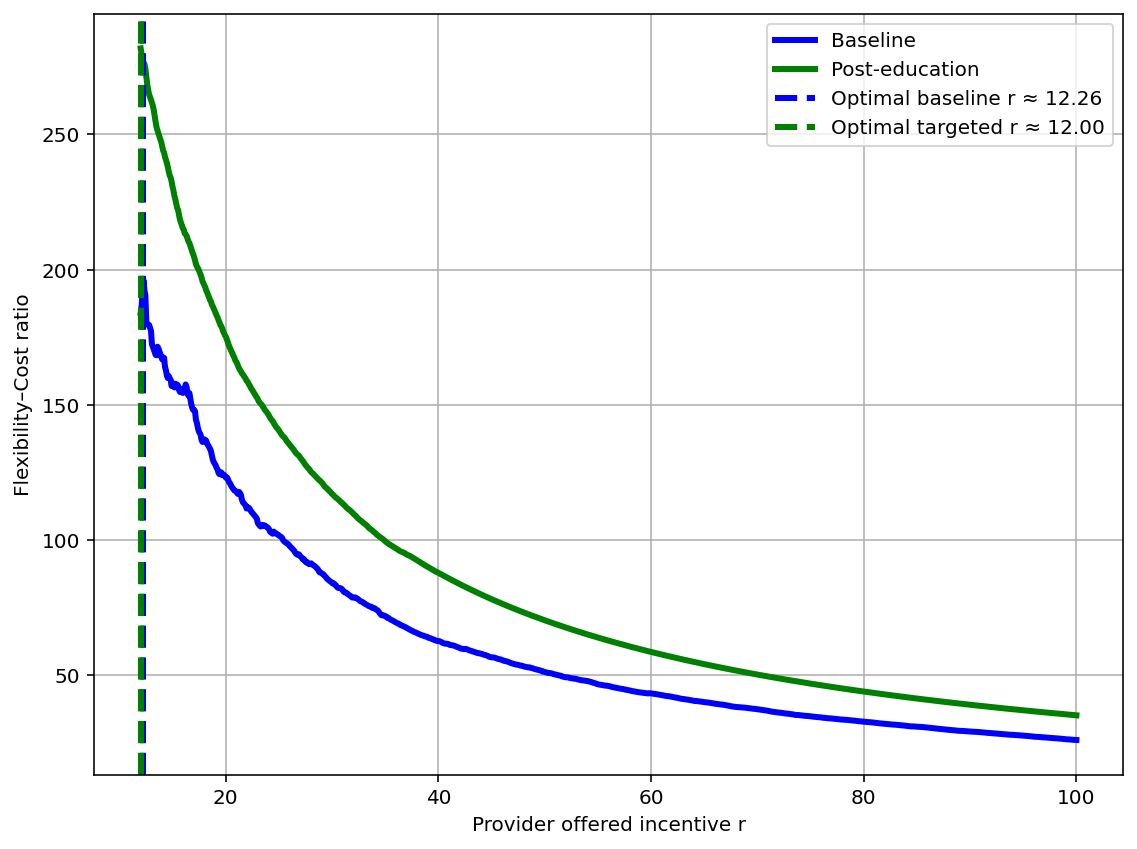} & \includegraphics[scale=0.35]{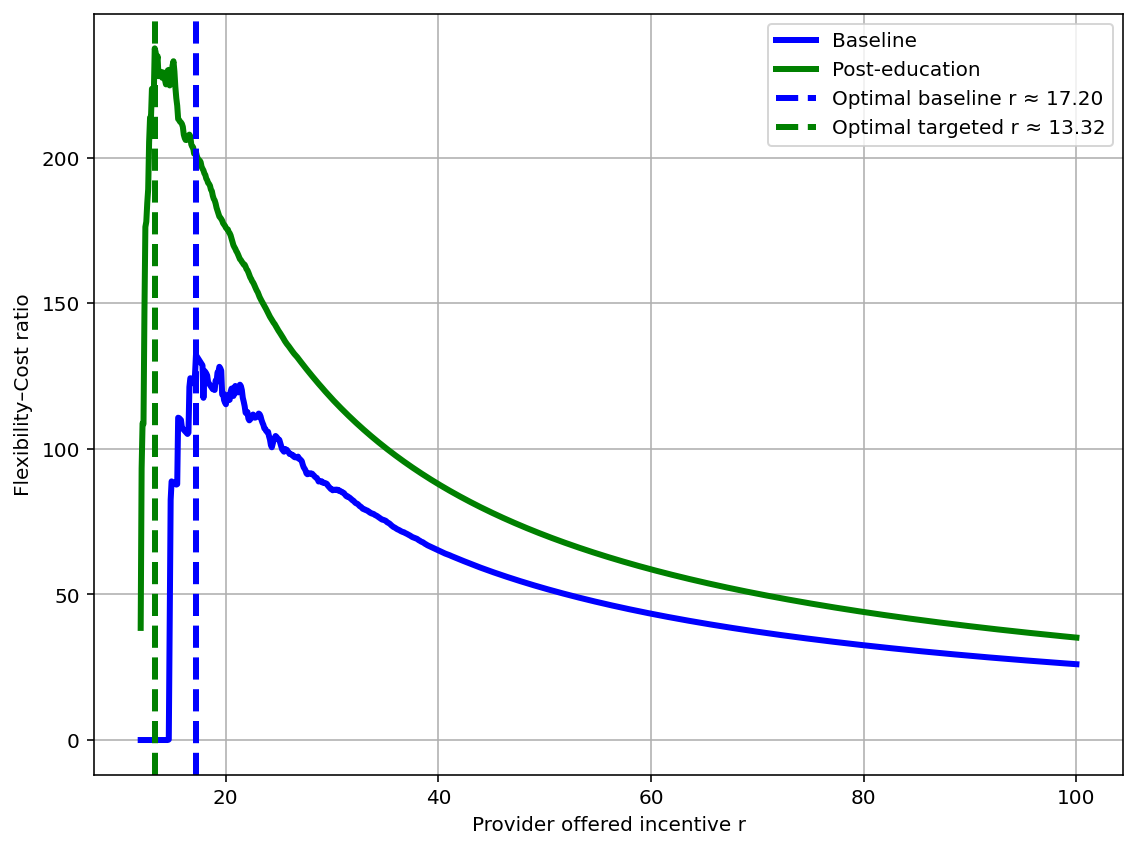}
     \end{array}$
    \caption{Flexibility-Cost ratio for the baseline offered incentives (blue) vs the educated offered incentives (green) sampled from $\mathcal{U}[a,b]$ (left) and from $\mathcal{N}(\mu, \sigma^2)$ (right).} 
    \label{fig:Educate reduce r min}
\end{figure}

Figure \ref{fig:Educate reduce r min} compares the Flexibility–Cost ratio before and after the ``education'' intervention, implementing \eqref{eq:educate}. We evaluate the education when baseline $r_{\min, n} \sim \mathcal{U}[10, 100]$, while the ``educated'' $r^{\text{target}}_{\min, n} \sim \mathcal{U}[5, 40]$, left panel in \ref{fig:Educate reduce r min}. Then we evaluate the education when baseline $r_{\min, n} \sim \mathcal{N}(30, 25)$, while the ``educated'' $r^{\text{target}}_{\min, n} \sim \mathcal{N}(40, 5)$, right panel in \ref{fig:Educate reduce r min}. Further, we take $\delta_n = 100$, $c_{\text{admin}} = 0.04$ and $s_n \sim \mathcal{U}[2, 5]$. Visually, in both plots, the green (educated) curve is shifted relative to the blue, indicating that the same offered incentive produces a different trade-off after education. Education further alters the mapping from offered incentives to realized flexibility in two ways. It can reduce required incentives by increasing intrinsic non-monetary benefits, and/or change QoE perception so that the same bitrate reduction yields different $\Delta U_n$. Practically, this means providers can often improve the optimal offered incentive, measuring marginal gains from education per user segment and targeting education where it both meaningfully lowers $r_{\min, n}$ and aligns with users whose bitrate reductions yield large flexibility.

\subsection{Learning from data}
\begin{figure}[t]
    \centering
    $\begin{array}{cc}
     \includegraphics[scale=0.35]{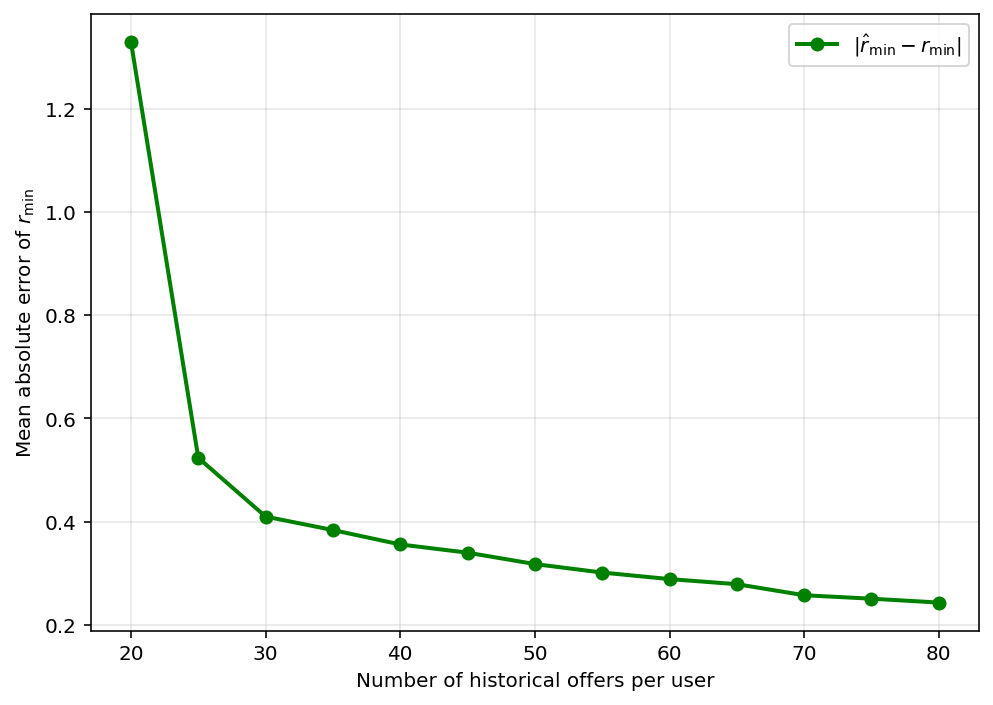} & \includegraphics[scale=0.35]{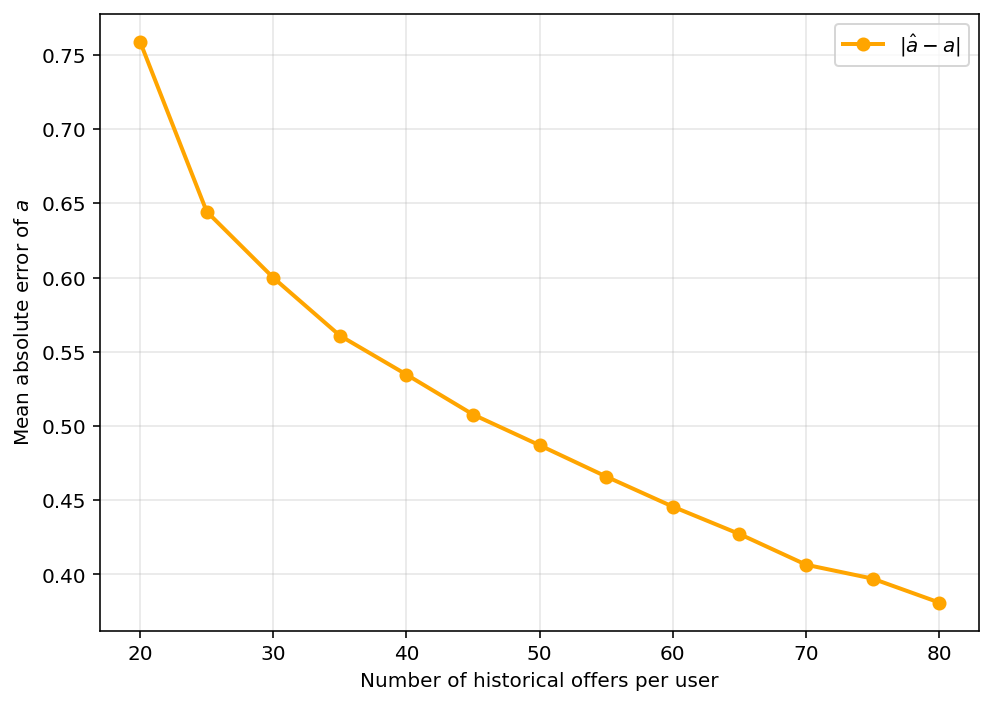}
    \end{array}$
    \caption{Mean error $|\widehat{r}_{\min, n} - r_{\min, n}|$ (left), and $|\widehat{\delta}_n - \delta_n|$ (right), for the uniform $\mathcal{U}[a, b]$.}
    \label{fig:Mean absolute error uniform}
\end{figure}

\begin{figure}[t]
    \centering
    $\begin{array}{cc}
     \includegraphics[scale=0.35]{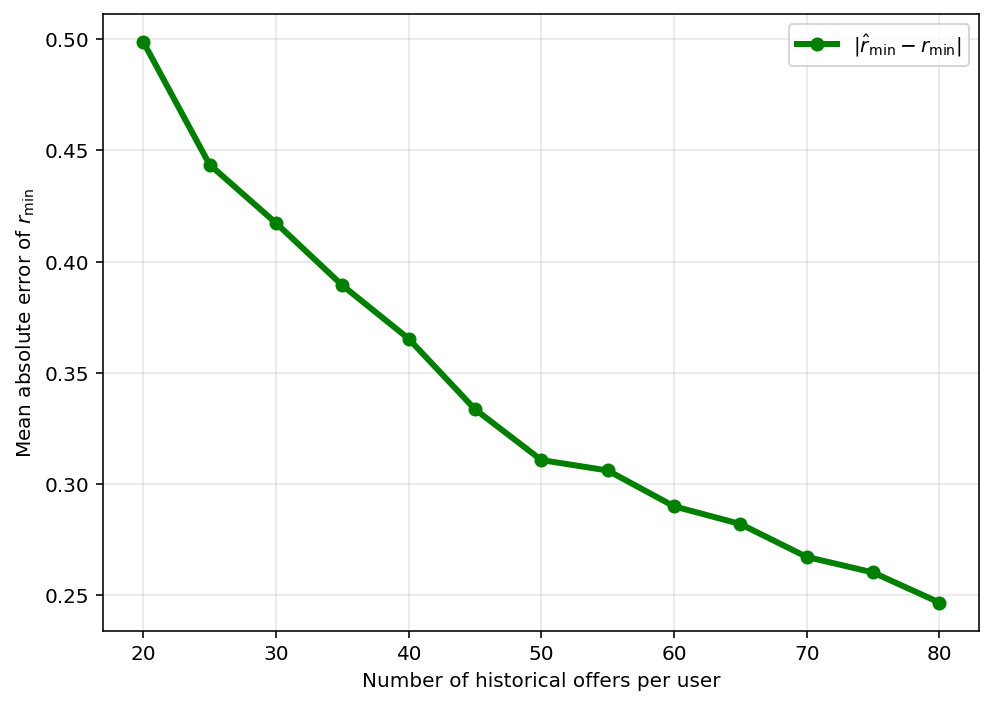} & \includegraphics[scale=0.35]{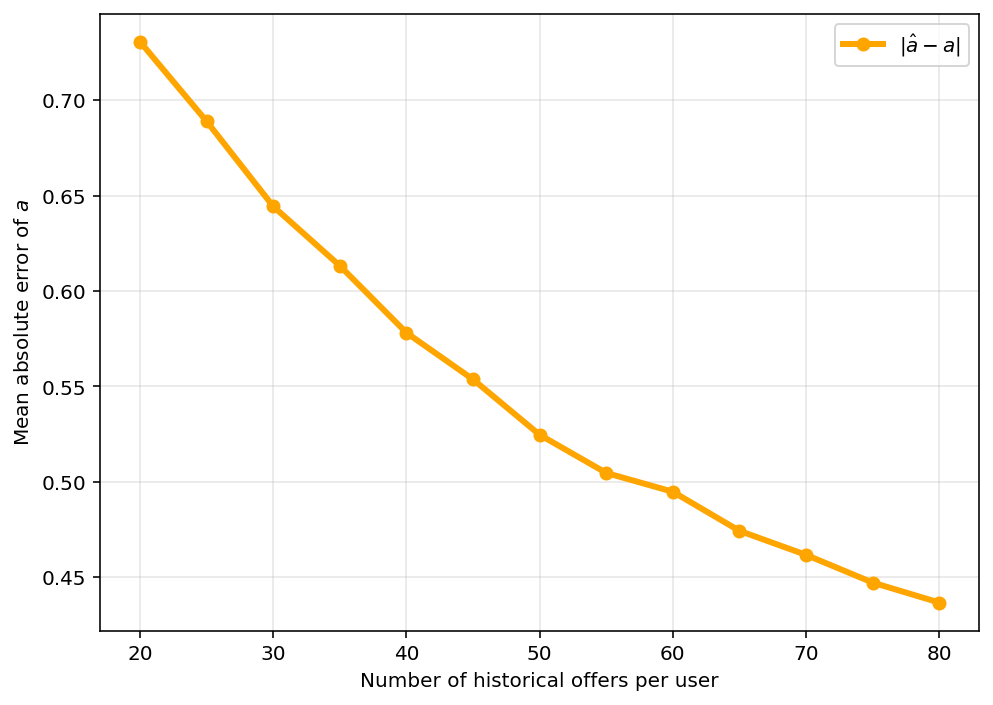}
    \end{array}$
    \caption{Mean error $|\widehat{r}_{\min, n} - r_{\min, n}|$ (left), and $|\widehat{\delta}_n - \delta_n|$ (right), for the normal $\mathcal{N}(\mu, \sigma^2)$.}
    \label{fig:Mean absolute error normal}
\end{figure}

In this setting, we evaluate the user acceptance model as presented in Subsection~\ref{subsec:learning from data}. Here, the provider takes into account up to $m$ past offers per user, with $m \leq 80$, along with users' decisions for each. Then it estimates the true values for the $r_{\min, n}$ and the $\delta_n$. In Figure~\ref{fig:Mean absolute error uniform}, $r_{\min, n} \sim \mathcal{U}[2, 8]$ and $\delta_n \sim \mathcal{U}[0.5, 3]$, and the provider offers incentives $r_n \sim \mathcal{U}[5, 4]$. We observe that as the provider agglomerates more historical data, its estimations improve monotonically. The same patter appears when $r_{\min, n} \sim \mathcal{N}(6, 1)$ and $\delta_n \sim \mathcal{N}(50, 9])$, and the provider offers incentives $r_{\min, n} \sim \mathcal{N}(0, 100)$, see Figure~\ref{fig:Mean absolute error normal}. Under standard regularity conditions (e.g., independent samples), as $m$ increases, the estimation errors tend to zero and the expected estimation error is asymptotically monotone decreasing, roughly proportional to $1/\sqrt{m}$, which supports the theoretical model in Subsection \ref{subsec:learning from data}.

\section{Conclusions}\label{sec:conclusions}

This work introduced a unified user acceptance model for incentivizing greener video streaming, integrating QoE, incentive thresholds, and altruism into a stochastic framework. Synthetic-data experiments revealed that higher environmental awareness does not guarantee lower incentive needs, while personalized offers significantly reduce provider costs compared to uniform policies. Our findings highlighted three practical insights for large-scale deployment. Firstly, personalized incentives enable cost-effective network flexibility; secondly, exposure to eco-friendly defaults lowers acceptance thresholds; and thirdly, learning of user responsiveness supports sustainability targets. Future research should validate these findings with real-world data and analyze the long-run dynamics of acceptance thresholds. Overall, the proposed framework lays the groundwork for scalable, data-informed approaches towards optimized, sustainable video streaming services.

\vspace{3pt}

\section*{Acknowledgment}

This work has been partly developed in the scope of the project EXIGENCE, which has received funding from the Smart Networks and Services Joint Undertaking (SNS JU) under the European Union (EU) Horizon Europe research and innovation programme under Grant Agreement No 101139120. Views and opinions expressed are however those of the author(s) only and do not necessarily reflect those of the EU or SNS JU.

%\printbibliography

 % \printbibliography
\bibliographystyle{plain} % or your preferred style
\bibliography{refs} % without the .bib extension

\end{document}